\documentclass[%
 aip,
 amsmath,amssymb,
 reprint,%
]{revtex4-1}
\usepackage[english]{babel}
\usepackage{graphicx}
\usepackage{dcolumn}
\usepackage{bm}
\usepackage{xcolor}
\usepackage[utf8]{inputenc}
\usepackage[T1]{fontenc}
\usepackage{mathptmx}
\usepackage{etoolbox}
\usepackage[hidelinks]{hyperref}
\makeatletter
\def\@email#1#2{%
 \endgroup
 \patchcmd{\titleblock@produce}
  {\frontmatter@RRAPformat}
  {\frontmatter@RRAPformat{\produce@RRAP{*#1\href{mailto:#2}{#2}}}\frontmatter@RRAPformat}
  {}{}
}%
\makeatother
\begin{document}

\preprint{AIP/123-QED}

\title[Explosive synchronization in networks of Type-I neurons with electrical synapses]{Explosive synchronization in networks of Type-I neurons with electrical synapses}

\author{Akshay S Harish}
\email[Corresponding author: ]{akshay.s.harish@gmail.com, p20200068@goa.bits-pilani.ac.in}
\affiliation{Department of Physics, BITS Pilani K K Birla Goa Campus, Zuarinagar, Goa 403726, India}

\author{Gaurav Dar}
\email{gdar@goa.bits-pilani.ac.in}
\affiliation{Department of Physics, BITS Pilani K K Birla Goa Campus, Zuarinagar, Goa 403726, India}


\begin{abstract}
Explosive synchronization (ES), which was observed in the scale-free network of the Kuramoto model [J. Gómez-Gardeñes et al.,Phys.Rev.Lett.106,128701(2011)], has been studied widely in the oscillator model. However, investigations of ES in neuronal networks, in spite of their importance in neuroscience, are limited and restricted to specific models. In this work, we explore the nature of the transition to synchronization in a class of neurons, namely Type-I neurons. Leveraging the mapping between networks of weakly heterogeneous Type-I neurons and the Kuramoto model [P.Clusella et al, Chaos 32, 013105 (2022)] under weak coupling, we investigate whether the conditions known to induce ES in the Kuramoto model also do so in Type-I neurons. The neurons are coupled through electrical synapses and placed on a scale-free and star networks with complete and partial degree-frequency correlation conditions. Our simulations show ES in networks of Quadratic Integrate and Fire (QIF) neurons, the normal form of Type-I neurons close to SNIC bifurcation, under weak heterogeneity. We further generalize this phenomenon to networks of Type-I Morris-Lecar neurons, under conditions similar to those of QIF neurons. Thus, this work suggests conditions under which ES can arise in Type-I neurons close to SNIC bifurcation under weak heterogeneity and weak electrical coupling.
\end{abstract}
\maketitle
\begin{quotation}
Explosive or abrupt transitions to synchronization have implications for several brain processes, such as epileptic seizure and switching between conscious states. Such transitions have been studied extensively in the abstract Kuramoto model. Although previous studies have reported explosive synchronization (ES) in neuronal networks, they are model-dependent and lack the generality to extend to other neuron models. Our work bridges this gap by demonstrating ES in a class of neurons. We achieve this by establishing ES in the normal form of Type-I neurons — the QIF model, and further affirming it in a Biophysical model exhibiting Type-I behavior. Leveraging the mapping between the Type-I neurons and the Kuramoto model \cite{clusella_kuramoto_2022}, we realize theoretical predictions of ES under different conditions on neuronal networks. The generality of these results across a class of neurons, together with their origin in the Kuramoto framework, takes us a step closer to understanding abrupt transitions in the brain. 
\end{quotation}
\section{Introduction}
Synchronization is a widely observed phenomenon across complex systems. Physical systems, such as coupled metronomes \cite{pantaleone_synchronization_2002}, Josephson junctions \cite{Wiesenfeld_synchronization_1996}, power-grid networks \cite{Motter2013}, fireflies \cite{mirollo_sync_1990, McCrea_A_Model_2022}, and neuronal networks \cite{ward_synchronous_2003}, all show varying degrees of synchronization. The onset of synchronization can occur smoothly, resembling a second order phase transition, or abruptly with accompanying hysteresis as in a first order transition. This abrupt transition is called  explosive synchronization (ES). 
Abrupt synchronization of neurons is hypothesized as the mechanism behind epileptic seizures \cite{jiruska_synchronization_2013} and also for the onset of Anesthetic-induced unconsciousness \cite{kim_functional_2016}.
Recent works hypothesize that the brain operates in a critical state close to a phase transition \cite{Fontenele_criticality_2019, serena_landau_2018, Khoshkhou_spike_2019}.  This emphasizes the need for a better understanding of synchronization in neuronal networks.\par
Explosive synchronization was first demonstrated in a scale-free network of Kuramoto oscillators with complete degree-frequency correlation \cite{gomez-gardenes_explosive_2011}.
Since then, there have been significant efforts in generalizing the conditions for ES in networks of Kuramoto oscillators and their variants. Relaxing the stringency of this first work, imposing only partial correlation \cite{pinto_explosive_2015} and adding disorder to the frequencies \cite{skardal-Disorder-2014} also exhibited ES on scale-free networks. Variants of the Kuramoto model with a phase lag can undergo ES on other network topologies if some extent of degree-frequency correlation exists \cite{kundu_synchronization_2019,kundu_transition_2017}.  Other studies employing the Kuramoto model have also examined alternate interplays between structures and dynamics, which include frequency weighted coupling \cite{zhang_explosive_2013}, adaptive coupling \cite{zhang_explosive_2015, Avalos_Emergent_2018}, and frequency-gap networks \cite{leyva_explosive_2013}. 
Multilayer networks of Kuramoto-Sakaguchi oscillators display ES for a wider class of network topologies, even without any structure-dynamics correlation \cite{kachhvah_multiplexing_2017, kachhvah_explosive_2021}. 
Beyond the Kuramoto model, degree-frequency correlated networks of Rössler oscillators and a few experimental systems - Electrical circuits \cite{leyva_explosive12}, mercury beating-oscillator \cite{kumar_experimental_2015}, and photochemical oscillators \cite{calugaru_first-order_2020} - show a sharp transition to synchronization. In several models, it is found that the nature of coupling is a key ingredient in inducing explosive synchronization \cite{SHARMA20192051, sharma_explosive_2021, ramesan_explosive_2022}. A more comprehensive account of the existing work on explosive synchronization can be found in this review \cite{bayani_explosive_2022}. \par
Despite the significance of ES to better understand brain processes, studies showing ES in neuronal networks are sparse. The neuron models studied exhibit ES under specific network topologies and with additional model-specific constraints. 
Izhikevich neurons and Chialvo neurons undergo explosive synchronization on a scale-free network without degree frequency correlation \cite{khoshkhou_beta-rhythm_2018, roy_assortativity-induced_2021}. Izhikevich neurons also exhibit such a sharp transition in the ER network. Small-world networks of Hodgkin-Huxley and Chialvo neurons also undergo an explosive synchronization \cite{khoshkhou_explosive_2020, boaretto_mechanism_2019}. The FHN model shows ES with nonlocal coupling \cite{verma_explosive_2022} and on a degree-frequency correlated scale-free network in the presence of noise \cite{chen_explosive_2013}. The nature of the transition - first order or second order - in neuronal networks can also depend on the type of synapses connecting the neurons \cite{khoshkhou_beta-rhythm_2018, khoshkhou_explosive_2020}.
The brain consists of a large variety of neurons, each requiring a specific model. The nature of transitions in these neuronal networks is model specific, limiting a broad understanding of synchronization transitions in the Brain network. To overcome this barrier, we explore explosive synchronization in a class of neurons. \par
For this work, we chose Type-I neurons, a class that is characterized by arbitrarily low frequencies near the onset of oscillations. Type-I neurons are predominantly found in the cortex and hippocampus \cite{Prescott_Biophysical_2008, Laing_2003}. Transition from rest to oscillations in these neurons occurs through the Saddle Node on Invariant Circle bifurcation. The Quadratic Integrate-Fire (QIF) model represents the normal form of all Type-I neurons close to the bifurcation point. Taking the QIF model as a prototype of all Type-I neurons, we explore explosive synchronization transition in networks of QIF neurons. To examine the generality of the results obtained, we further analyze synchronization transitions for networks of Morris-Lecar (ML) neurons, known to display Type-I dynamics \cite{TSUMOTO2006293}.  This is a biophysical model originally developed to model the dynamics of barnacle muscle fiber \cite{MORRIS1981193}.\par 
Mean firing rate approaches based on dimensionality reduction have been successfully employed to analyze bifurcations and synchronization transitions in heterogeneous networks of QIF neurons \cite{laing_exact_2015, Pietras_exact_2019, Laing_effects_2021, pietras_2024_heterogeneous}. While some works \cite{laing_exact_2015, Pietras_exact_2019} report SNIC and subcritical Hopf bifurcations in the mean-field equations, these bifurcations alone do not guarantee explosive synchronization, as bistability—and the associated hysteresis—between synchronous and asynchronous states is also required. The derivation of low-dimensional mean-field equations for QIF populations commonly relies on Lorentzian heterogeneity, which allows for analytical tractability, but also imposes restrictions on the forms of heterogeneity that can be studied. In this work, we instead rely on an alternate analytical formulation of QIF networks. In the formulation, a network of electrically coupled QIF neurons with weak heterogeneity and weak electrical coupling map to the Kuramoto model with zero phase lag \cite{clusella_kuramoto_2022}. Building on the well documented studies of explosive synchronization in the Kuramoto oscillator network under various conditions, we explore the emergence of ES in Type-I neuronal networks. We carry out numerical simulations of these neuronal networks with varying degrees of correlation between degree and frequency. In particular, focusing on degree-frequency correlated scale-free and star networks of Kuramoto oscillators, as discussed in Gomez-Gardenes et al. \cite{gomez-gardenes_explosive_2011}, along with partially correlated and uncorrelated scale-free networks studied by Pinto et. al., \cite{pinto_explosive_2015}. \par
This paper is organized as follows. In section \ref{model} we describe the neuron models used in this work with details of the models, commenting on the nature of the bifurcations in these models. In sections \ref{qif_results} and \ref{ML_results}, results of numerical simulations carried out for networks of QIF neurons and ML neurons are presented, respectively. Additionally, in section \ref{qif_results}, we numerically validate the mapping between the QIF and the Kuramoto model. Section \ref{conclusion} gives the concluding remarks of the results discussed in the previous sections. 
\section{Model}\label{model}
In Type-I neurons, the change from excitatory to oscillatory state is achieved by the injected current and occurs through a saddle Node on invariant circle (SNIC) bifurcation. The Quadratic Integrate-and-Fire (QIF) model is a simple yet powerful neuronal model that captures the essential dynamics of Type-I neurons. The following equations describe the QIF model:
\begin{subequations}
\label{qif-main}
\begin{align} 
\tau \dot{V_i} = V_i^2 + \eta_i + \varepsilon I_{i, syn}(t)  \label{qif-main1} \\
 if \ V_i > V_p, \ then  \ \ V_i \rightarrow V_r \label{qif-main2}
\end{align}
\end{subequations}
where $V_i$ is the voltage of $i^{th}$ neuron in the network. The term $\eta_i$ represents the external current injected into the neuron, and the term $I_{i, syn}$ represents the net synaptic current flowing into neuron \textit{i}. When $V_i$ reaches a large threshold value, $V_p$ (often approximated as $+\infty$), it is reset to a large negative value, $V_r$ (often taken as $-\infty$), to mimic the action potential firing and reset process of a neuron. Eq.~(\ref{qif-main1}) describes the approach of a QIF neuron to the threshold by the normal form of SNIC bifurcation, while  Eq.~(\ref{qif-main2}) represents the reset process. When $\eta_i <0$, the neuron is in an excitable state, and if $\eta_i>0$, it oscillates, mimicking a train of action potentials (spikes). The frequency of spiking increases gradually with $\eta_i>0$ from an arbitrarily low value - a characteristic of Type-I neurons. A well known variable transformed version of the QIF model is the ``$\Theta$-neuron model" \cite{ermentrout_parabolic_1986}. This is a popular choice when the system is amenable to Ott-Antonsen ansatz and thereby to dimensionality reduction \cite{chandra_modeling_2017, so_networks_2014}. However, when interactions between the neurons are mediated by electrical synapses, synaptic coupling needs to be approximated to avoid singularity  \cite{ermentrout_gap_2006} in the $\Theta$-neuron model. \par
In our work, we couple the QIF neurons through electrical synapses, leading to the following form for Eq.~(\ref{qif-main1}).
\begin{equation}
	\tau \dot{V_i} = V_i^2 + \eta_i + \varepsilon g \sum_{j=1}^{N} A_{ij} (V_j - V_i)
    \label{qif-main-2}
\end{equation}
while the reset condition remains the same. In the above equation, g is the strength of the electrical synapse, $A$ is the adjacency matrix which characterizes the network topology with $A_{ij}=1$ if $i^{th}$ neuron is connected to the $j^{th}$ neuron, and $A_{ij}=0$ if there is no connection between them. There exists no self-connection, i.e., $A_{ii}=0$ for all $i$. We consider two network topologies - scale-free (SF) and star networks. The SF network has a degree distribution given by the power law $p(k) = k^{-\gamma}$ where the degree $k_i = \sum_j^{N} A_{ij}$ of neuron $i$ is the number of neurons in the network connected to it. In a star network, the single hub neuron is connected to all the other neurons, which are not themselves interconnected. An SF network is characterized by hub-like structures. Hence, star networks act as a precursor to the SF network. We consider networks of QIF neurons with varying degrees of correlation between their degrees and the frequencies.\par
Between two successive spikes, voltage dynamics of a free-running QIF neuron is governed by Eq.~(\ref{qif-main-2}) with $g=0$. Its solution \cite{pfeuty_electrical_2003}, starting from the voltage value $ V_r$ at $t=0$ is 
\begin{equation}
	V_i(t) = \sqrt{\eta_i} \tan \Bigg[ \frac{t \sqrt{\eta_i}}{\tau} + \arctan \Bigg( \frac{V_r}{\sqrt{\eta_i}} \Bigg) \Bigg]
    \label{qif-soln}
\end{equation}
with its firing frequency $\Omega_i = {(\pi \sqrt{\eta_i})}/{(\tau \arctan({V_p} / {\sqrt{\eta_i}}))}$. The phase $ \theta_i   = \Omega_i \ t$ of a QIF neuron can be expressed in terms of its voltage as  
\begin{equation}
    \label{finite_reset_phase}
	\theta_i  =  \bigg{[} \frac{\pi }{ \arctan(\frac{V_p}{\sqrt{\eta_i}})}) \bigg{]}   \bigg{[} \arctan(\frac{V_i}{\sqrt{\eta_i}}) - \arctan(\frac{V_r}{\sqrt{\eta_i}})\bigg{]}
\end{equation}
Accordingly,  $\theta_i \rightarrow 2\pi $ as  $V_i(t) \rightarrow V_p$ and  $\theta_i \rightarrow 0 $ as $V_i(t) \rightarrow V_r$. To quantify the degree of synchronization, we use an instantaneous order parameter $r(t) = \frac{1}{N} | \sum_{j=1}^{N} e^{i \theta_j(t)} | $, which is time averaged over steady states to yield a global order parameter $ R = \langle r(t) \rangle$. R varies from $0$ to $1$, with $R=1$ representing the completely synchronized state and $R=0$ the asynchronous state of the network. \par
For a network of electrically connected QIF neurons with $\eta_i = \bar{\eta} + \varepsilon \chi_i$ taking $\bar{\eta} >0$ and $\chi_i>0$, heterogeneity is weak if $\varepsilon \chi_i << \bar{\eta}$. Likewise, $\frac{\varepsilon g}{\bar{\eta}} << 1 $ keeps the coupling between neurons weak. In a recent work \cite{clusella_kuramoto_2022} with $V_p = -V_r = \infty$, and assuming weak heterogeneity and weak coupling, a network of electrically coupled QIF neurons shown to be reduced to the Kuramoto model with zero phase lag \cite{clusella_kuramoto_2022} expressed as
\begin{equation}
 \dot{\theta_i} =  \bigg{(}\frac{2\sqrt{\bar{\eta}}}{\tau} + \frac{\varepsilon \chi_i}{\tau \sqrt{\bar{\eta}}}\bigg{)} +  \frac{\varepsilon g}{\tau} \sum_j^{N} \sin (\theta_j - \theta_i) 
 \label{reducedKM}
\end{equation}
where $\omega_i =(\frac{2\sqrt{\bar{\eta}}}{\tau} + \frac{\varepsilon \chi_i}{\tau \sqrt{\bar{\eta}}})$ and $\lambda = \varepsilon g/\tau $ are the frequency and the coupling strength respectively of this reduced QIF-Kuramoto model. \par
The Morris-Lecar model is a conductance based biophysical model formulated to describe the dynamics of the giant barnacle muscle fibre. The following equations constitute the dynamics of a Morris-Lecar neuron connected to other neurons in the network through electrical synapses.
\begin{subequations}
\label{ML_main}
\begin{eqnarray}
        C_M \frac{dV_i}{dt} =&& - {g_L} (V_i - V_l) - g_{Ca} M_{\infty} (V_i - V_{Ca}) -g_K W_i (V_i - V_K) \nonumber\\
        &&  + I_{i,ext} - g_{syn} 
        \sum_{j=1}^{N}A_{ij}( V_j - V_i) \label{ML_main_a} \\
         \frac{dW_i}{dt} = &&\frac{W_{\infty } - W_i}{\tau_{W}}
         \label{ML_main_b}
\end{eqnarray} 
\end{subequations}
In Eq.~(\ref{ML_main_b}), $W_i$ represents the recovery variable of the $i^{th}$ neuron. $V_l$, $V_{Ca}$, $V_K$ represents the equilibrium potentials of the leak, $Ca^{2+}$, and $K^+$ currents respectively. $g_l$, $g_{Ca}$, and $g_K$ are the maximum conductance of the corresponding ionic channels. In our work, we have chosen $V_l= -60 mV$, $V_{Ca}= 120 mV$, $V_K= -80 mV$, $g_l = 2 \ mS/cm^2$, $g_{Ca} = 4.0 \ mS/cm^2$, and $g_K = 8 \ mS/cm^2$. The capacitance of the neuron membrane, $C_M$ is taken as $20 \mu F/Cm^2$. 
$M_{\infty}(V_i) = 0.5[1+ \tanh{ \{(V_i-V_1)/V_2 \} } ]$ and $W_{\infty}(V_i) = 0.5[1+ \tanh{ \{(V_i-V_3)/V_4} \} ] $ are the steady-state activation function of the $Ca^{2+}$ and $K^+$ currents respectively and  $\tau_w (V_i) = 1/ [\phi \cosh{ \{ (V_i -V_3)/2V_4 \} } ]$ is the time constant of the $K^+$ activation, with $V_1 = -1.2 mV$, $V_2 = 18.0 mV $, $V_3= 12.0 mV$, and $V_4= 17.4 mV$. These parameter values, chosen from \cite{TSUMOTO2006293}, ensure that the neurons are operating in the Type-I regime. 
External current, $I_{i, ext}$, drives the neurons from an excitable to an oscillatory state through SNIC bifurcation. In an oscillatory state, the free-running frequency of an uncoupled neuron increases with the external current. We consider the same network topologies as those for QIF neurons. Degree-frequency correlation is imposed by fixing $I_{i, ext} = I_0 + \varepsilon k_i$. The value of $I_0$ is chosen to keep the ML neuron close to the bifurcation point so that its dynamics can be approximated by the QIF model. Conductance-based models of a neuron, like the Morris-Lecar, do not admit an analytical expression for their phase. Hence, for such a periodically spiking neuron, we compute the phase numerically, which increases linearly between successive spikes as
\begin{equation}
    \phi_i(t) = 2 \pi \frac{t - T_{n,i}}{T_{n+1, i} - T_{n,i}},  \  \  T_{n,i} \leq t < T_{n+1,i}
    \label{phase}
\end{equation}
where $T_{n,i}$ and $T_{n+1, i}$ are the $n^{th}$ and $(n+1)^{th}$ spike times respectively of the $i^{th}$ neuron. 
The global order parameter computed using these phases ($\phi_i$) gives the degree of synchrony in the network.

\section{Results}
In this section, we discuss the results of numerical simulations carried out for electrically coupled networks of QIF and ML neurons with varying magnitudes of correlation between degree and frequency. We begin by assessing the validity of the reduction from the QIF network to the reduced QIF–Kuramoto model.
\subsection{Networks of QIF neurons} \label{qif_results}
The dynamics of QIF neurons are governed by Eq.~(\ref{qif-main-2}) along with the reset rule of Eq.~(\ref{qif-main2}).  We fix $V_p = - V_r = 750$ and $\tau=1$ for all neurons. 
\begin{figure}
\centering
\includegraphics[width = \linewidth]{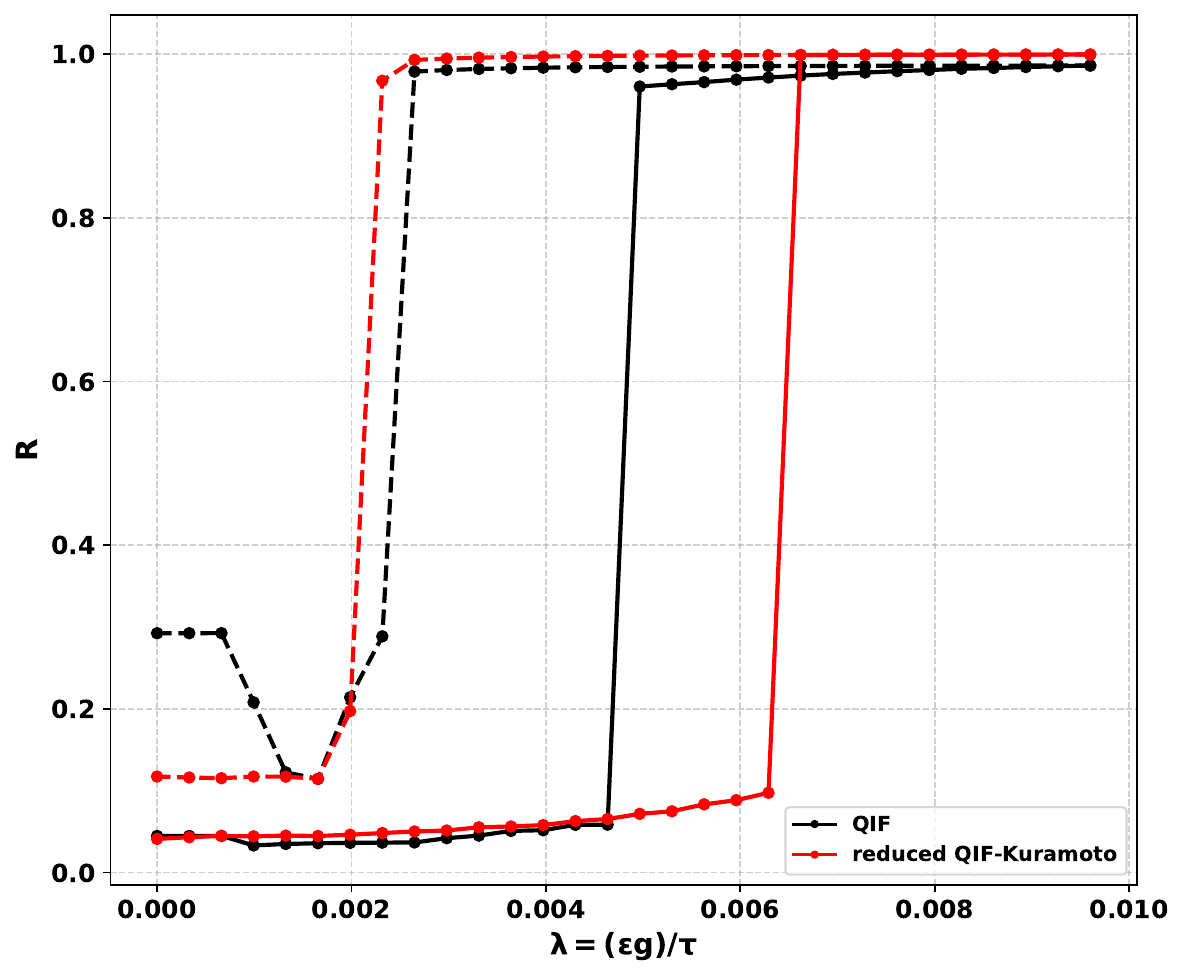}
\caption{comparison of a degree-frequency correlated star network of 31 QIF neurons and the corresponding reduced QIF-Kuramoto model. The solid lines represent the forward g simulation, while the dotted line represents
the backward simulation. For these simulations $\bar{\eta} = 20$, $\varepsilon = 0.008$. }
\label{figure1}
\end{figure}
Synaptic strength $g$ is identical between all connected neurons. Simulations are performed into steady states for increasing values of $g$, starting at $g = 0$, to a maximum $g$ value. Thereafter, simulations are similarly carried out for decreasing values of $g$. Final steady state voltages for the current $g$ are used as the initial conditions for the next $g$ value. At $g=0$, we start with uniformly distributed voltage values. The order parameter is calculated in the steady state for every value of $g$ in forward and backward runs.  Since for a QIF neuron, $\eta$ is proportional to its frequency\cite{ermentrout_parabolic_1986}, degree-frequency correlation is imposed following the rule  $\eta_i = \bar{\eta} + \varepsilon k_i$. We choose $\bar{\eta} > 0$ to place neurons in the oscillatory regime. We take $\bar{\eta} = 20 $ and  $\varepsilon \approx 10^{-2} \in (0.005, 0.01)$, ensuring weak heterogeneity. In this regime, we numerically assess the validity of mapping the QIF model to the reduced QIF-Kuramoto model. In \cite{clusella_kuramoto_2022}, comparisons between the two models were made with zero heterogeneity. \par
\begin{figure}
\centering
\includegraphics[width = \linewidth]{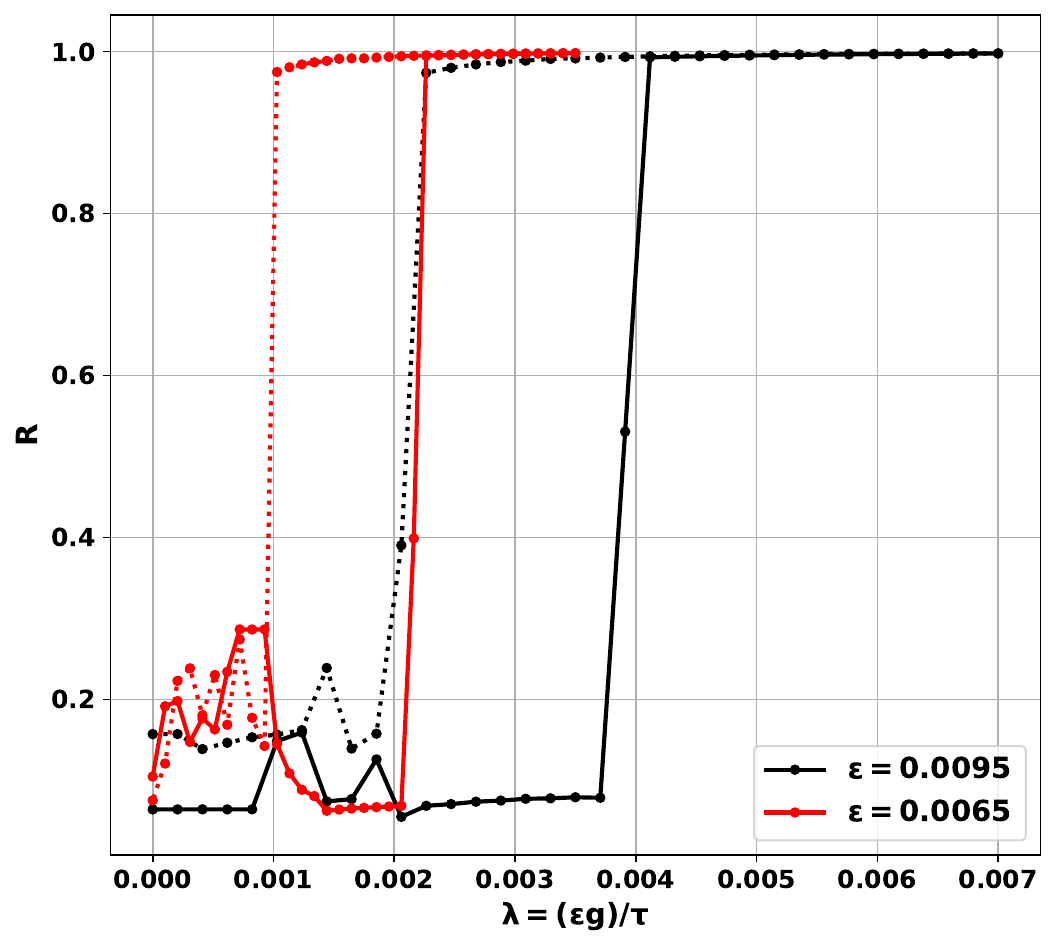}
\caption{Effect of $\varepsilon$ on the transition points in a 21 complete degree-frequency correlated star networks of QIF neurons. The solid lines represent the forward g simulation, while the dotted line represents the backward simulation. Other parameters are the same as in Fig.~\ref{figure1}. }
\label{figure2}
\end{figure} 
We first carry out simulations of a degree-frequency correlated star network of QIF neurons (Eq.~(\ref{qif-main-2})) and the corresponding reduced QIF-Kuramoto network (Eq.~(\ref{reducedKM})). Both the networks are solved using the RK4 \cite{note1} 
method with step size $dt =2.5\times 10^{-4}$ and the synchronization transitions are compared.
The phases in Eq.~(\ref{finite_reset_phase}) is a good approximation to phase in Eq.~(\ref{reducedKM}) with the choice of $V_p = -V_r = 750$ (see Appendix \ref{appA}). Thus, the order parameters calculated from these two are comparable. Simulations are carried out till $T_f = 50000$, and the order parameter is calculated over the final 30000 time units.   
In these simulations, when $g$ values are less than a certain value, a small randomness is added \cite{leyva_local_2025} to the $\eta$ values during the backward run. This is done to slightly perturb the system out of the synchronized branch. 
\begin{figure}
\centering
\includegraphics[width = \linewidth]{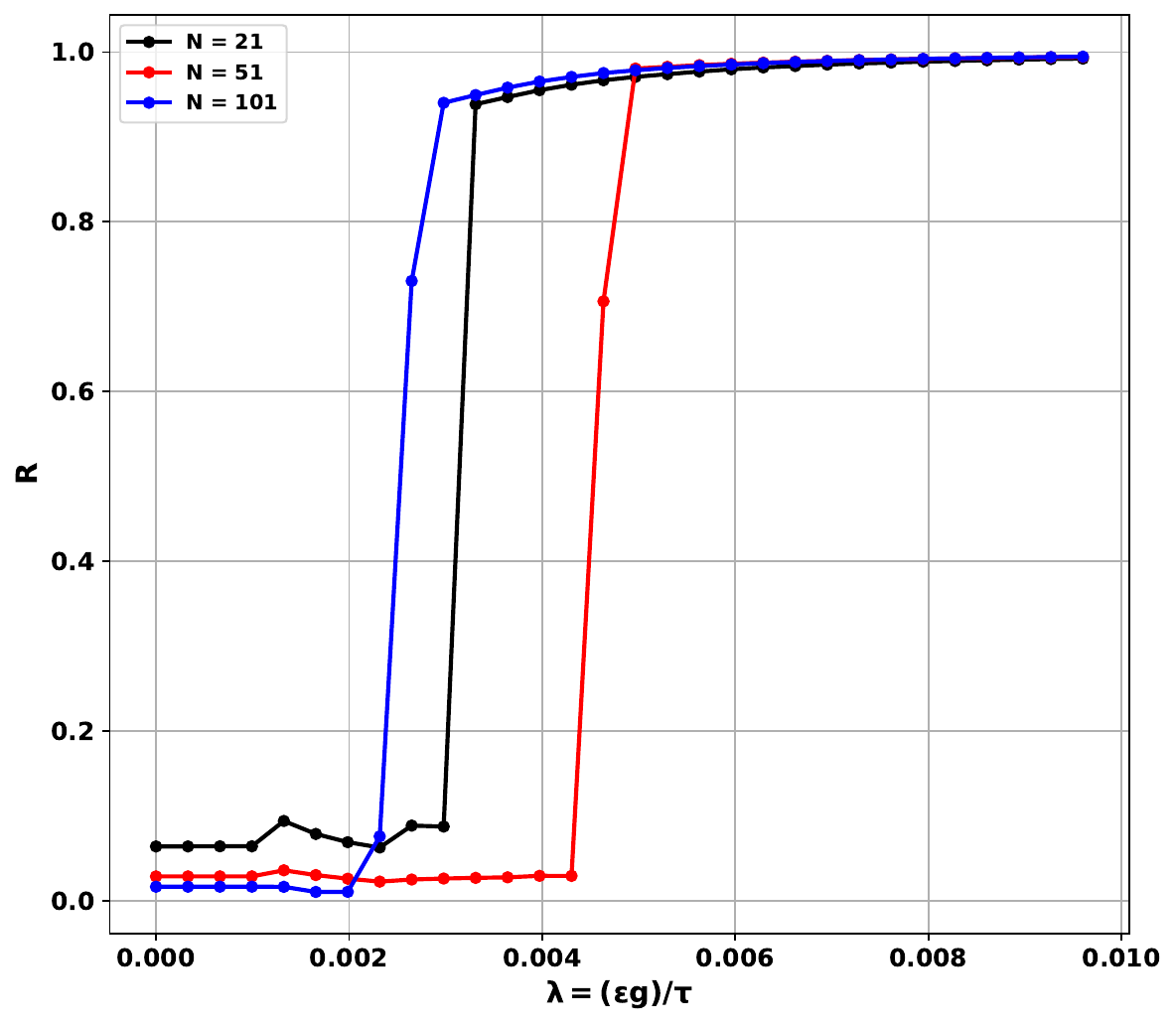}
\caption{Forward simulations of Star networks of QIF neurons with complete degree-frequency correlation. N is the number of neurons in the network. For these simulations $\bar{\eta} = 20$, $\varepsilon = 0.008$. }
\label{figure3}
\end{figure}
\begin{figure*}
\centering
\includegraphics[width = \linewidth]{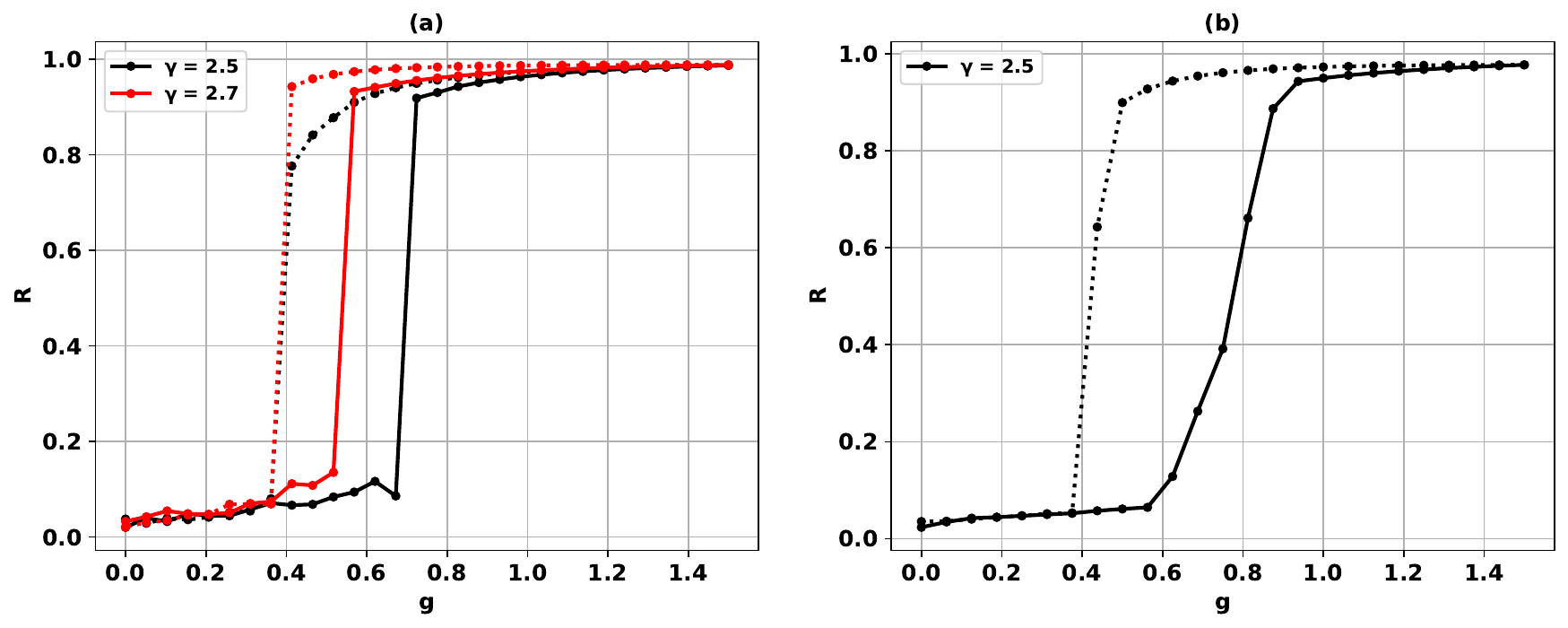}
\caption{ A sharp transition to synchronization accompanied by hysteresis in the order parameter is seen for scale-free networks of 1000 QIF neurons with complete degree-frequency correlation. (a)  $\gamma = 2.4$ (black line) and $\gamma =2.7$ (red line). (b) average of order parameter for 50 realizations of $\gamma = 2.5$ SF network. In all cases $\bar{\eta} = 20 $, $\varepsilon = 0.01$ with solid curves representing forward simulations and dotted lines representing backward simulations.}
\label{figure4}
\end{figure*}
Fig.~\ref{figure1} shows an abrupt transition in the order parameter accompanied by hysteresis, depicting explosive synchronization in the star network of degree-frequency correlated QIF neurons. The backward transition point of QIF networks shows excellent quantitative agreement with that of the corresponding reduced QIF-Kuramoto model. The forward transition points of the two models lie close to one another but exhibit sensitivity to the $\varepsilon$ values (not included in the figure), which is likely due to finite-size effects.\par
Further, we study the effect of heterogeneity on the backward transition points of star networks of QIF neurons. The mean-field calculation for a fully degree-frequency correlated star network of Kuramoto oscillators yields the critical coupling strength $\lambda_c$ for the backward transition below which a stable synchronized state does not exist \cite{gomez-gardenes_explosive_2011}. The corresponding theoretical $\lambda_c$ for the star network of QIF neurons is calculated in Appendix~\ref{appB}. This gives 
\begin{equation}
    \lambda_c = \frac{\varepsilon g_c}{\tau} = \frac{\varepsilon (K-1)}{\tau\bar{\eta}(K+1)}
    \label{critical_g}
\end{equation}
For $\varepsilon= 0.0095$ and $\varepsilon= 0.0065$, theoretically predicted $\lambda_c$ values are $\approx 0.0019$ and $\approx 0.0013$ respectively. In Fig.~\ref{figure2}, it is seen that the backward transitions occur at $\lambda$'s close to these values predicted by the above equation. The above results validated the mapping between the Kuramoto model and the weakly heterogeneous QIF model with weak electrical coupling. We also performed simulations of QIF star networks of different sizes. Fig.~\ref{figure3} shows forward simulation results. There is a sharp transition to synchronization across different sizes of star networks. Each of these is accompanied by hysteresis (not shown in the figure). \par
\begin{figure*}
\centering 
\includegraphics[width = 1.0\linewidth]{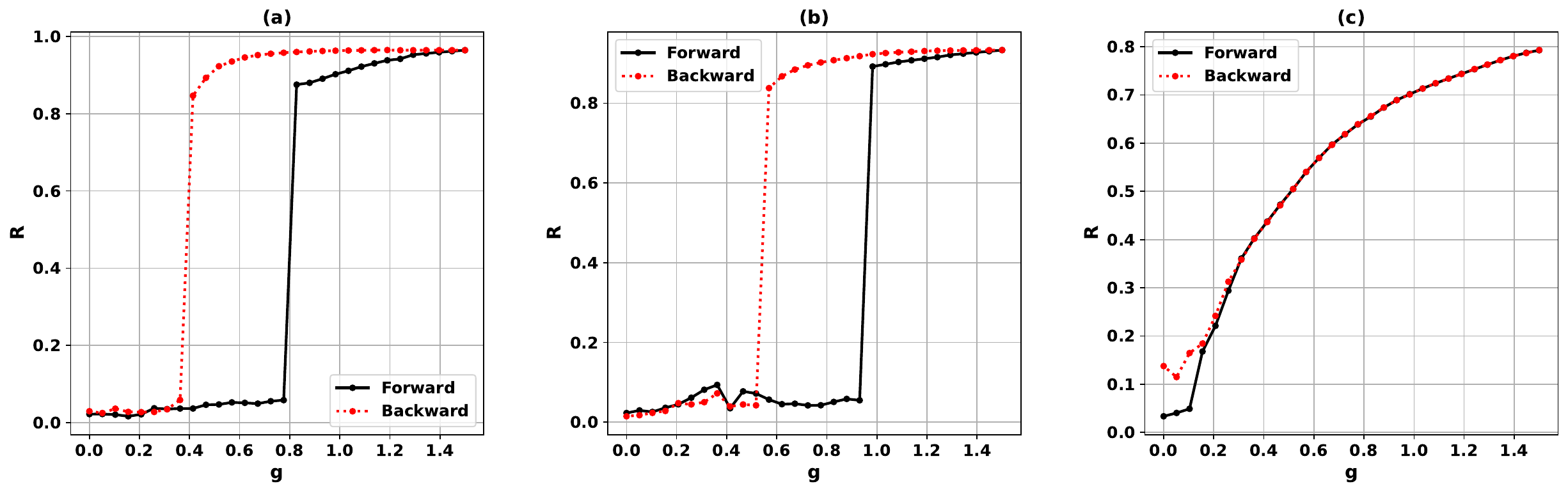}
\caption{Plots of 1000 QIF neurons connected by an SF network with varying degree-frequency correlations. In all cases, the same SF network with $\gamma = 2.3$ is used. (a) Complete degree-frequency correlation. (b) Only the top $10 \%$ of neurons with the highest degree are assigned frequencies correlated with their degree, while the rest are assigned random frequencies. (c) All neurons are assigned random frequencies. In both (b) and (c), random frequencies are drawn from a Lorentzian distribution $p(k) = a/\pi((k-k_0)^2 - a^2) $ with $k_0 = 0$ and $a=1$. For all cases $\bar{\eta} = 20 $ and $\varepsilon = 0.01$. }
\label{figure5}
\end{figure*}
Further, we examine other QIF network conditions that are known to be sufficient to induce ES in the Kuramoto model. In particular, scale-free networks of Kuramoto oscillators with varying degrees of correlation between degree and frequency.
\begin{figure}
\centering
\includegraphics[width = \linewidth]{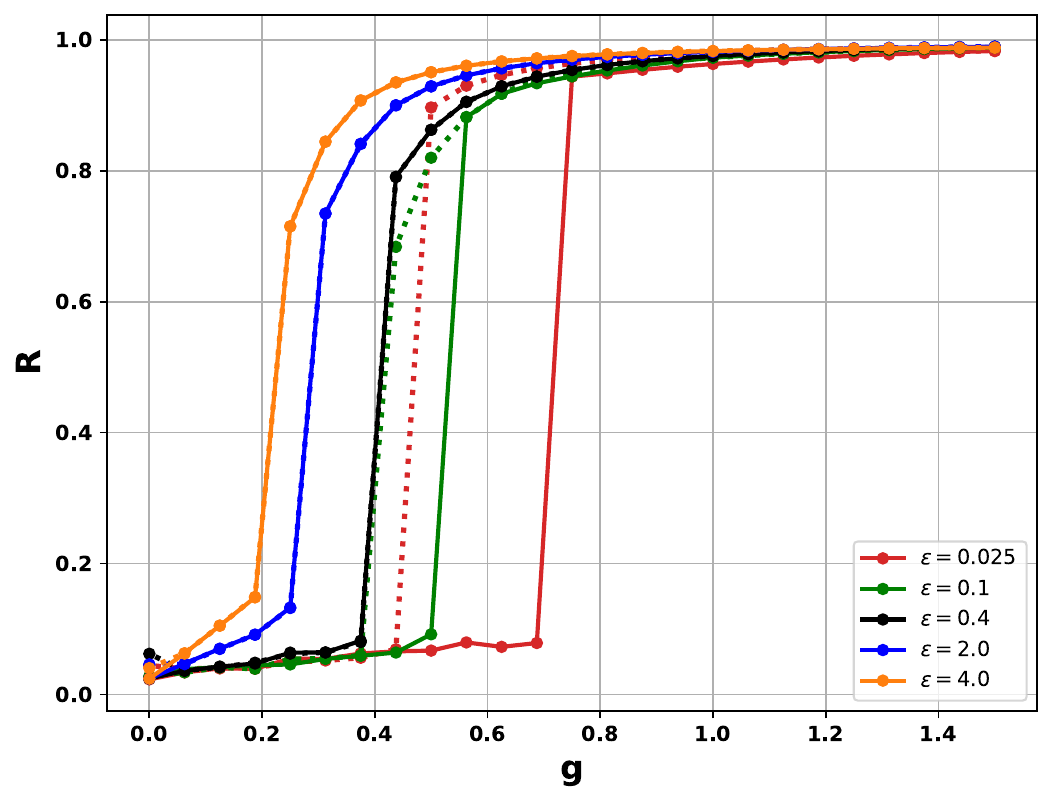}
\caption{As the heterogeneity ($\varepsilon$) is increased, hysteresis width decreases for the SF networks of 1000 degree-frequency correlated QIF neurons and the transitions become less sharp. In the figure, the solid lines represent the forward simulation, and the dotted lines represent the backward simulation. All the other parameters are considered the same as those taken in Fig.~\ref{figure4}a, $\gamma=2.5$ case.}
\label{figure6}
\end{figure}
We first simulate 1000 fully degree-frequency correlated QIF neurons arranged on SF networks. The SF networks are built via a configuration model. This model gives control over the degree exponent $\gamma$ of the degree distribution $p(k)$. We have considered  $2 < \gamma < 3$ in this work. For the degree-frequency correlated scale-free networks in our simulations, we take $\bar{\eta}=20$ and $\varepsilon = 0.01$. With these parameters, $99 \%$ of the neurons have $20.0 \leq \eta_i \leq 20.5 $, maintaining the weak heterogeneity.
We solve Eq.~(\ref{qif-main-2}) using Euler method with a small step size $dt$ ($\approx 10^{-4}$) over the duration $T_f = 10000$. The first $ 5000 $ time units are discarded as transients, and the rest are used for evaluating $R$. Simulations are performed by varying g following the same method as described for star networks. Variations in the order parameter while increasing and decreasing values of $g$ are shown in Fig.~\ref{figure4}. A sharp transition to synchronization, along with hysteresis, is clearly seen - exhibiting an explosive synchronization. This result is robust to the degree exponent of the scale-free network, as shown in Fig.~\ref{figure4} for different $\gamma$ values. We carried out simulations over 50 realizations of the SF network with the same $\gamma$. Hysteresis and sharp transitions persist in all of them. Forward transition point is found to be at $g_{fc}$  $ = 0.766 \pm 0.092$ and the backward transition point at $g_{bc} =  0.453 \pm 0.027$. Fig.~\ref{figure4}b shows the averaged values of $R$ over these 50 realizations. The Kuramoto model with zero phase lag is known to undergo explosive synchronization even with partial degree-frequency correlation \cite{pinto_explosive_2015}. Exploiting the mapping between the QIF neuronal network and the Kuramoto model \cite{clusella_kuramoto_2022}, we also investigated synchronization transition in partially degree-frequency correlated QIF networks. To achieve partial degree-frequency correlation in the QIF network, a fraction of the neuron population is assigned frequencies following the $\eta_i = \bar{\eta} + \varepsilon k_i$ rule. The rest are assigned $\chi_i$ randomly from a Lorentzian distribution, making the frequencies random. 
\begin{figure}
\centering
\includegraphics[width = \linewidth]{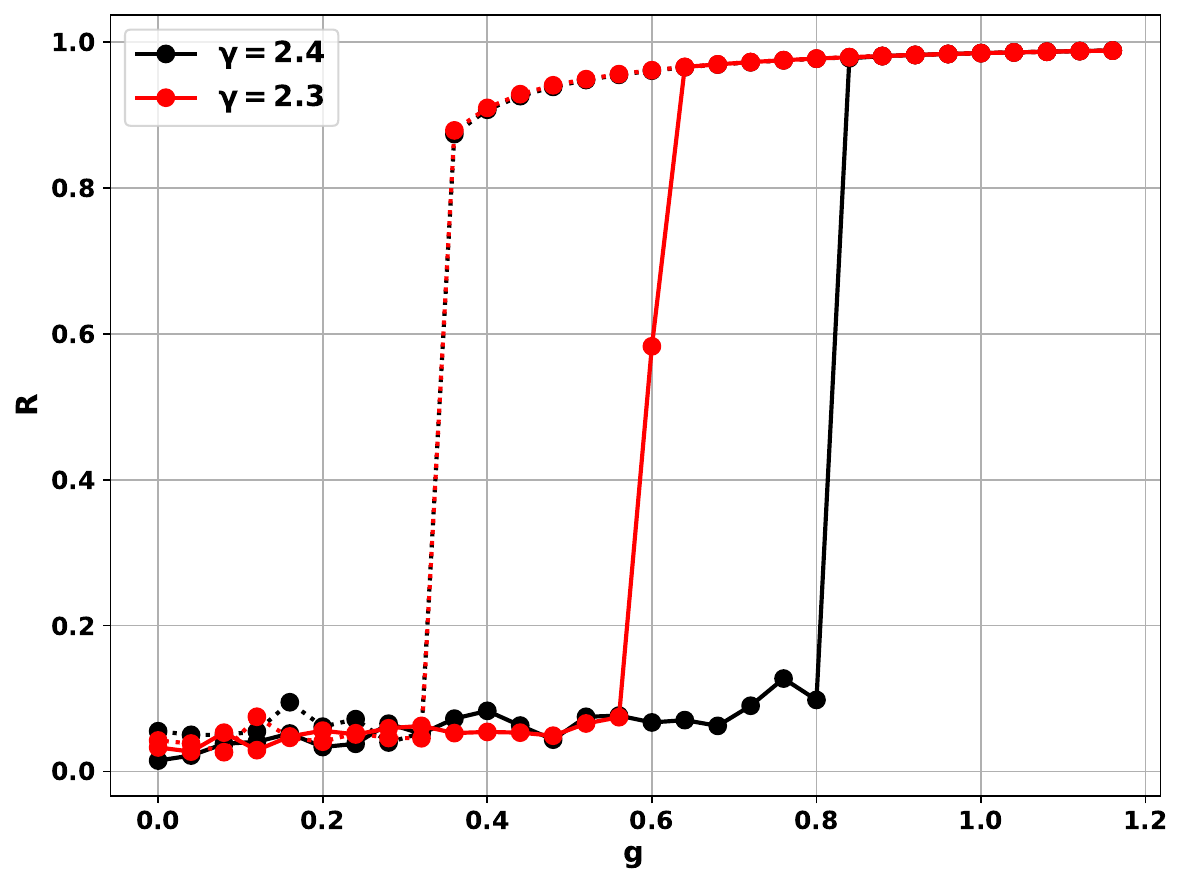}
\caption{ Explosive synchronization along with Hysteresis in the order parameter for a scale-free network of 1000 Degree-frequency correlated ML neurons.$\gamma = 2.4$ (black line) and $\gamma =2.3$ (red line). For both cases, $I_{0} = 43.0 $, $\varepsilon = 0.01$, and the solid curves represent the forward simulations while the dotted lines represent the backward simulations.}
\label{figure7}
\end{figure}
\begin{figure*}
\centering
\includegraphics[width = \linewidth]{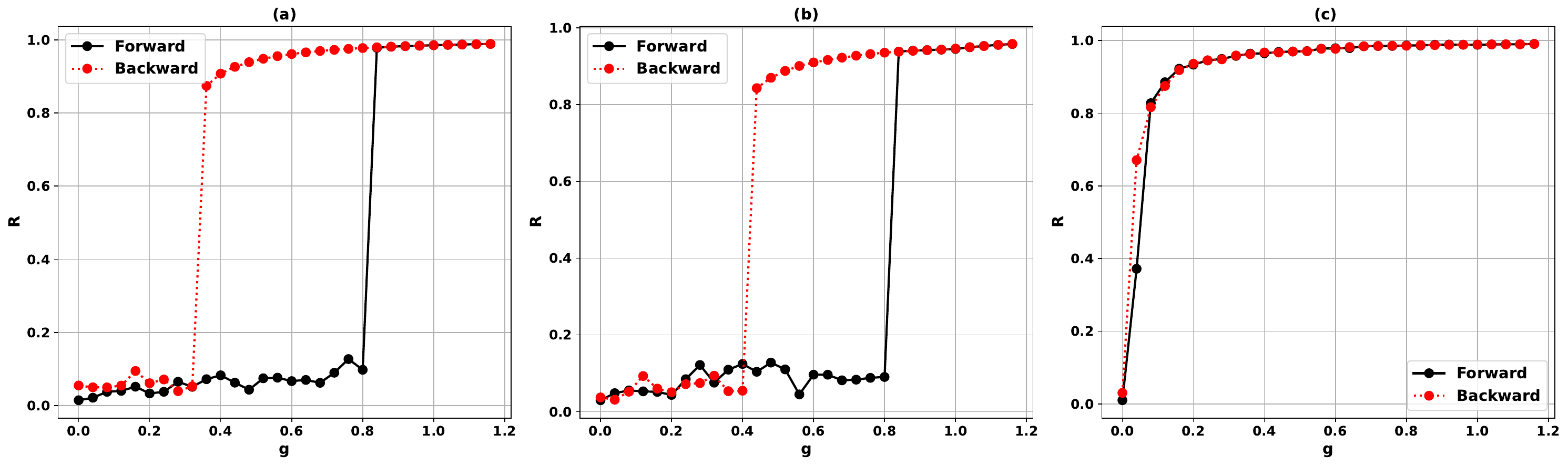}
\caption{Plots of 1000 ML neurons connected through an SF network. In all cases, the same SF network with $\gamma = 2.4$ is used. (a) Complete degree-frequency correlation. (b) Only the top $10 \%$ of neurons with the highest degree are assigned frequencies correlated with their degree, while the rest are assigned random frequencies. (c) All neurons are assigned random frequencies. In both (b) and (c), random frequencies are drawn from a Lorentzian distribution $p(k) = a/\pi((k-k_0)^2 - a^2) $ with $k_0 = 0$ and $a=1$. For all cases $I_0 = 43.0 \mu A /cm^2 $ and $\varepsilon = 0.01$.}
\label{figure8}
\end{figure*}
Fig.~\ref{figure5} shows that the first-order transition to synchronization observed in a fully degree-frequency correlated SF network of QIF neurons persists even when only $10 \%$ of the highest degree neurons are correlated. In contrast,  when the frequencies of all the neurons are uncorrelated with their degrees, the transition becomes continuous.\newline
$\ $ We examine the effect of increasing the level of heterogeneity. As shown in Fig.~\ref{figure6}, for a degree-frequency correlated SF network of QIF neurons hysteresis width progressively shrinks as $\varepsilon$ is increased and becomes negligible as $\varepsilon$ approaches 0.4. On further increasing $\varepsilon$, the sharpness in the transition decreases and tends to become continuous.
In contrast, for the reduced QIF-Kuramoto model, the hysteresis width is unaffected by $\varepsilon$ (not shown). Thus, for sufficiently large heterogeneity, ES conditions for the Kuramoto model may not serve as a reliable guide for selecting ES conditions for the QIF model. 

\subsection{Networks of Morris-Lecar neurons} \label{ML_results}
In the previous section, we reported that the SF networks of the normal form of Type-I neurons (the QIF model) show explosive synchronization with complete and partial degree-frequency correlation. However, the transition becomes smooth if the degrees and the frequencies are uncorrelated. Now, we generalize these results to Type-I Morris-Lecar neurons operating close to the SNIC bifurcation on the same network topologies with different levels of degree-frequency correlation. A Morris-Lecar neuron, modeled by Eq.~(\ref{ML_main}), shows Type-I behavior for the chosen parameter values of section \ref{model}. With these parameter values an SNIC bifurcation, separating excitable and oscillatory states, occurs at $I_{ext} \approx 39.6 \mu A /cm^2 \ $ \cite{TSUMOTO2006293}.
Following section \ref{model}, we correlate the frequency of a neuron with its degree as per the rule $I_{i, ext} = I_0 + \varepsilon k_i$, with $I_0 = 43.0 \mu A /cm^2$ and $\varepsilon = 0.01$. With these parameters, $I_{ext}$ for nearly all neurons lies between $43.0 \mu A /cm^2$ and $44.0 \mu A /cm^2$ for scale-free networks considered in our simulations (see Appendix \ref{appC} for details), consistent with weak heterogeneity. Moreover, with this range of $I_{ext}$, all neurons lie close to the SNIC bifurcation. The  Simulations were performed for scale-free networks of 1000 ML neurons. The strength of the electrical synapse was varied following the procedure considered for scale-free networks of QIF neurons. The order parameter was calculated using the phases defined in Eq.~(\ref{phase}). \par
First, we consider complete degree-frequency correlation with different $\gamma$ values of the scale-free network. Results are shown in Fig.~\ref{figure7}. The transition to synchronization is explosive with a sharp jump in the order parameter accompanied by hysteresis.

Partial degree-frequency correlation is achieved by assigning frequencies to the uncorrelated neurons in the same way as was done for the QIF networks. Simulation results are shown in Fig.~\ref{figure8}. We observe an abrupt transition to a synchronized state with only $10 \%$ of correlation. When the neuron frequencies are completely uncorrelated to their degrees, the transition becomes continuous. With these results of ML neuronal networks, we show that any Type-I neurons, with weak heterogeneity and coupling, can show ES if the neuronal network satisfies the conditions needed for ES in the corresponding Kuramoto model. \par
These observations verify the mapping among these three models, which approximates the ML model with the QIF model, and the QIF model with the Kuramoto model.

\section{Conclusion} \label{conclusion}
We have demonstrated the emergence of explosive synchronization in scale-free and star networks of Type-I neurons with weak electrical coupling and weak intrinsic heterogeneity. Specifically, we have shown the occurrence of ES in the normal form of Type-I neurons (QIF model) and generalized it in a biophysical model of Type-I neurons (Morris-Lecar model) close to the SNIC bifurcation. Our results in Fig.~\ref{figure4} and \ref{figure7}  show that ES is robust to the degree exponent of scale-free networks. Fig.~\ref{figure5} and \ref{figure8} show that ES is observed for both complete and partial degree-frequency correlation, and the transition becomes continuous when there is no correlation. Thus, the explosive synchronization observed here is network-induced rather than a manifestation of the macroscopic dynamics observed through mean firing rate models.\par
The mapping between the QIF and the Kuramoto model \cite{clusella_kuramoto_2022} served as a motivation for employing ES scenarios of the Kuramoto networks to our work. To validate this mapping between weakly heterogeneous and weakly coupled QIF and the Kuramoto model, we compared the simulation and analytical results of the reduced QIF-Kuramoto model with the corresponding degree-frequency correlated star network of QIF neurons as shown in Fig.~\ref{figure1} and \ref{figure2}. The quantitative agreement was close, specifically for the critical coupling of backward transitions. At strong heterogeneity, the mapping does not appear to hold as seen in our simulations of SF networks of QIF neurons with increasing heterogeneity. As shown in Fig.~\ref{figure6}, hysteresis vanishes, and the transition tends to become continuous as heterogeneity increases. \par
Our findings broaden the presence of ES, showing that it is not limited to specific neuronal models but can arise across a broad class of neurons. ES is hypothesized as the mechanism for epileptic seizures and the onset of anesthetic-induced unconsciousness. Understanding the conditions of explosive synchronization in neuronal networks could shed light on normal and pathological brain dynamics. Leveraging the mapping between Type-I neurons and the Kuramoto model, a large set of scenarios can be explored for explosive synchronization. This work is the first step towards it.

\section*{Data Availability Statement}
Data sharing is not applicable to this article as no new data were created or analyzed in this study.   

\appendix
\section{Choosing Threshold voltages ($V_p$) for the QIF model}
\label{appA}
\begin{figure}
\centering
\includegraphics[width = \linewidth]{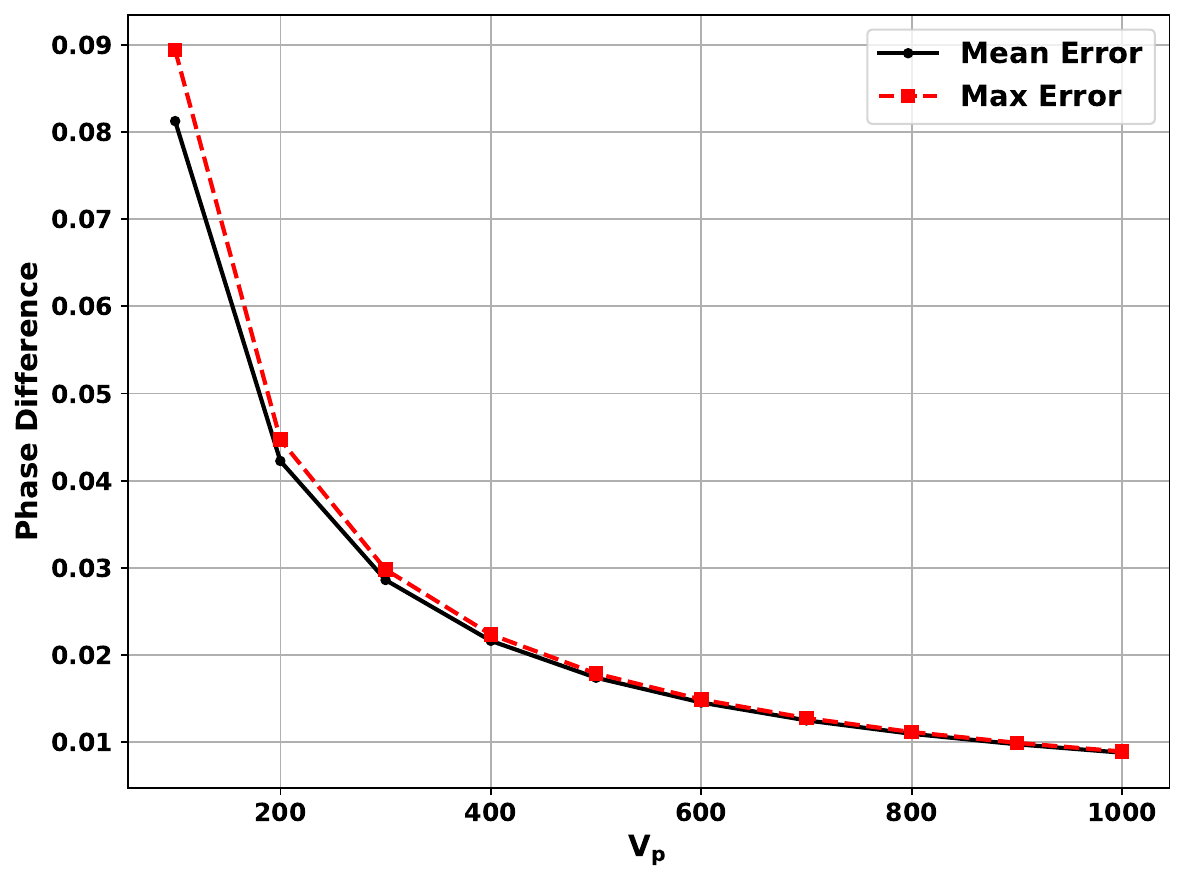}
\caption{ Increasing $V_p$ values decreases the error between the phase in Eq.~(\ref{finite_reset_phase}) and the asymptotic phase. Here $\eta = 20$ is taken in Eq.~(\ref{finite_reset_phase})}
\label{figure9}
\end{figure}

In our simulations, phases of QIF neurons are computed using Eq.~(\ref{finite_reset_phase}), while Eq.~(\ref{reducedKM}) employs phases computed in the asymptotic limit of $V_p \rightarrow \infty $. In this appendix, we show that the two phases approach each other as the threshold voltage is increased in Eq.~(\ref{finite_reset_phase}). 
We select a $V_p$ and consider 5000 uniformly distributed voltages between $V_p$ and $V_r$ (with $V_r = -V_p$). At these voltage values, we calculate the two phases and compute the mean and absolute differences. Fig.~\ref{figure9} shows that as the threshold voltage is increased, the phase of Eq.~(\ref{finite_reset_phase}) tend towards the asymptotic phase indicated by the decreasing mean and the maximum error.
This serves as a reference for selecting $V_p = 750$ in our numerical simulations. 
\section{Analytical expression for the critical coupling of degree-frequency correlated star network of QIF neurons}
\label{appB}
The mean-field analysis of the Kuramoto model on a star network with K leaves results in a critical coupling 
\begin{equation}
    \lambda_c = (\Omega-\omega_l)
    \label{appB1}
\end{equation}
where the synchronized state loses stability \cite{gomez-gardenes_explosive_2011}. Here, $\Omega$ and $\omega_l$ are the average frequency of the network and the common frequency of the leaves, respectively. Eq.~(\ref{appB1}) is obtained by considering phases in a frame rotating with frequency $\Omega$ and imposing the phase-lock conditions on the hubs and leaves. Furthermore, for a degree-frequency correlated star network with $\omega_l =1$ and $\omega_h =K$ as the frequencies of the leaves and hubs, respectively, it has been shown that \cite{gomez-gardenes_explosive_2011} Eq.~(\ref{appB1}) reduces to
\begin{equation}
    \lambda_c = \frac{K-1}{K+1}
\end{equation}
For the degree-frequency correlation considered in our work in the reduced QIF-Kuramoto model, Eq.~(\ref{reducedKM}), frequencies of the hub and leaves are 
\begin{subequations}
\begin{align}
\omega_h =\frac{2\sqrt{\bar{\eta}}}{\tau} + \frac{\varepsilon K}{\tau \sqrt{\bar{\eta}}} 
\label{appb3a} \\
\omega_l =\frac{2\sqrt{\bar{\eta}}}{\tau} + \frac{\varepsilon }{\tau \sqrt{\bar{\eta}}}
\label{appb3b}
\end{align}
\end{subequations}
The resulting average frequency is 
\begin{equation}
    \Omega = (2\sqrt{\bar{\eta}}/\tau) + (2 \varepsilon K/ (\tau \bar{\eta}(K+1)))
    \label{appB4}
\end{equation}
Substituting the frequencies  Eq.~(\ref{appB4}) and Eq.~(\ref{appb3b}) into Eq.~(\ref{appB1}) gives the critical coupling of the reduced QIF-Kuramoto model as 
\begin{equation}
    \lambda_c = \frac{\varepsilon g_c}{\tau} = \frac{\varepsilon (K-1)}{\tau\bar{\eta}(K+1)}
\end{equation}
This is the theoretical $\lambda_c$ for the star network of the reduced QIF-Kuramoto model, as mentioned in Eq.~(\ref{critical_g}).
\begin{figure}
\centering
\includegraphics[width = \linewidth]{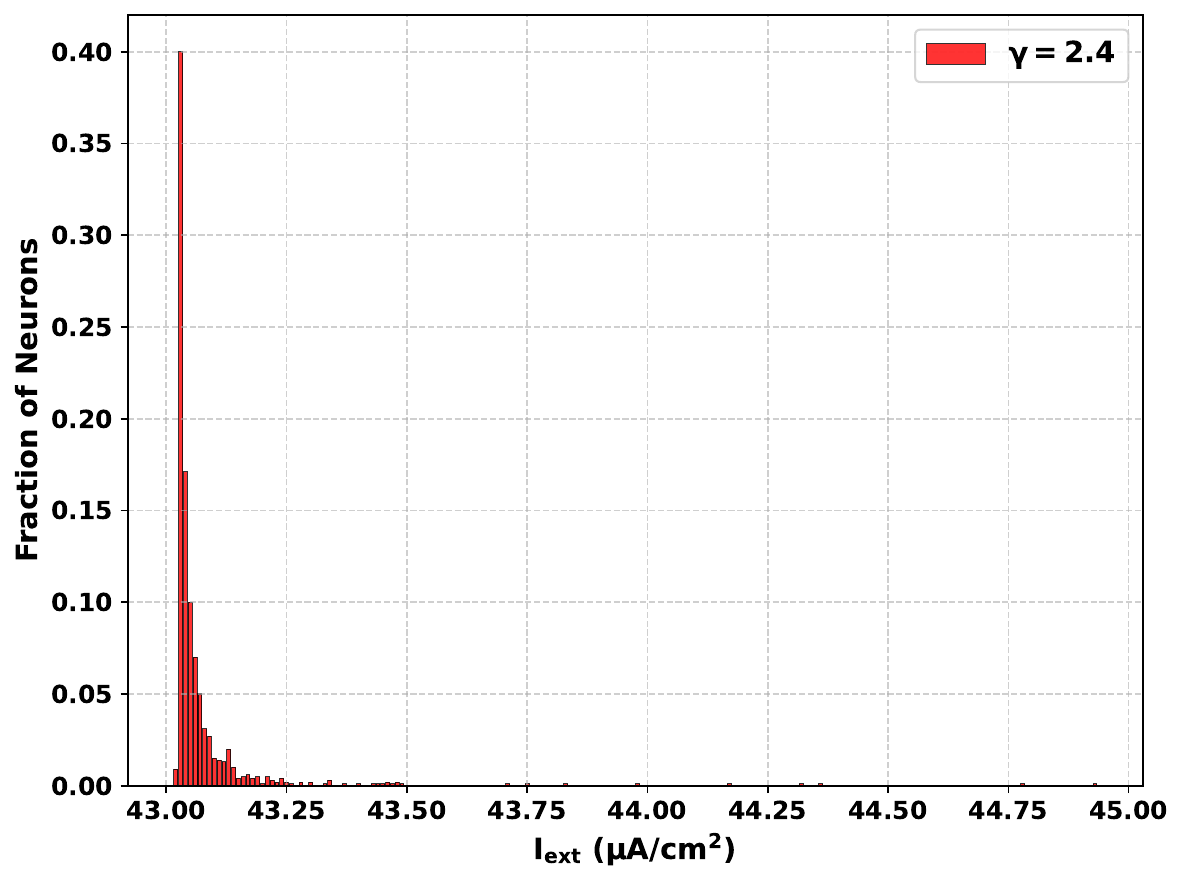}
\caption{Distribution of external current of degree-frequency correlated Morris-Lecar neurons with scale-free network ($\gamma = 2.4 $) considered in Fig.~\ref{figure8}. $99\%$ of the neuron fraction has $I_{ext}$ between 43.0 and 44.0 }
\label{figure10}
\end{figure}
\section{$I_{ext,i}$ distribution in Morris-Lecar scale-free neuronal network }
\label{appC}

We consider the scale-free network with $\gamma=2.4$ (used in Fig.~\ref{figure8}) and plot the distribution of external current $I_{ext, i}$ when the neurons are fully degree-frequency correlated following the $I_{ext,i} = I_0 + \varepsilon k_i$  rule. Fig.~\ref{figure10} shows that the external current distribution is peaked very close to the $I_0$ value, and $99 \%$ of the neuron population has $ 43.0 \mu A /cm^2 < I_{ext} < 44.0 \mu A /cm^2$. Therefore, in our simulations, neurons' intrinsic dynamics are weakly heterogeneous.


\begin{thebibliography}{53}
\bibitem{note1}{This higher order method is chosen over the Euler method to reduce the finite-size effect in the smaller star networks.}
\end{thebibliography}


\end{document}